\documentclass[]{spie}  %>>> use for US letter paper
\usepackage[]{graphicx}

\title{A charge transfer inefficiency correction model for the Chandra Advanced CCD Imaging Spectrometer} 

\author{C. E. Grant, M. W. Bautz, S. M. Kissel, and B. LaMarr
\skiplinehalf
Center for Space Research, Massachusetts Institute of Technology, \\
Cambridge, Massachusetts 02139
}

\authorinfo{Further author information: (Send correspondence to C.E.G.)\\C.E.G.: E-mail: cgrant@space.mit.edu}

\begin{document} 
  \maketitle 

%%%%%%%%%%%%%%%%%%%%%%%%%%%%%%%%%%%%%%%%%%%%%%%%%%%%%%%%%%%%%

\begin{abstract}
Soon after launch, the Advanced CCD Imaging Spectrometer (ACIS), one of the focal plane instruments on the Chandra X-ray Observatory, suffered radiation damage from exposure to soft protons during passages through the Earth's radiation belts.  The primary effect of the damage was to increase the charge transfer inefficiency (CTI) of the eight front illuminated CCDs by more than two orders of magnitude.  The ACIS instrument team is continuing to study the properties of the damage with an emphasis on developing techniques to mitigate CTI and spectral resolution degradation.  We will discuss the characteristics of the damage, the detector and the particle background and how they conspire to degrade the instrument performance.  We have developed a model for ACIS CTI which can be used to correct each event and regain some of the lost performance.  The correction uses a map of the electron trap distribution, a parameterization of the energy dependent charge loss and the fraction of the lost charge re-emitted into the trailing pixel to correct the pixels in the event island.  This model has been implemented in the standard Chandra data processing pipeline.  Some of the correction algorithm was inspired by the earlier work on ACIS CTI correction by Townsley and collaborators\cite{Townsley00,Townsley02}.  The details of the CTI model and how each parameter improves performance will be discussed, as well as the limitations and the possibilities for future improvement.

\end{abstract}


\section{INTRODUCTION}
\label{sect:intro}
The Chandra X-ray Observatory, the third of NASA's great observatories in space, was launched just past midnight on July 23, 1999, aboard the space shuttle {\it Columbia}\cite{cha2}.  After a series of orbital maneuvers Chandra reached its final, highly elliptical, orbit.  Chandra's orbit, with a perigee of 10,000~km, an apogee of 140,000~km and an initial inclination of 28.5$^\circ$, transits a wide range of particle environments, from the radiation belts at closest approach through the magnetosphere and magnetopause and past the bow shock into the solar wind.

The Advanced CCD Imaging Spectrometer (ACIS), one of two focal plane science instruments on Chandra, utilizes charge-coupled devices (CCDs) of two types, front- and back-illuminated (FI and BI).  Soon after launch it was discovered that the FI CCDs had suffered radiation damage from exposure to soft protons scattered off the Observatory's grazing-incidence optics during passages through the Earth's radiation belts\cite{gyp00}.  The BI CCDs were unaffected.  Since mid-September 1999, ACIS has been protected during radiation belt passages and there is an ongoing effort to prevent further damage and to develop hardware and software strategies to mitigate the effects of charge transfer inefficiency on data analysis.  One such strategy, post-facto correction of event pulseheights, is described here.  This correction algorithm has been implemented in the standard Chandra data processing pipeline.

This paper begins by defining the calibration data used to test correction algorithms in Section~\ref{sect:data}.  Section~\ref{sect:raddamage} describes the characteristics of ACIS radiation damage.  The CTI correction algorithm is outlined in Section~\ref{sect:cticorr}.  The performance of the correction algorithm is presented in Section~\ref{sect:perf} and further discussed in Section~\ref{sect:disc}.

\section{DATA} 
\label{sect:data}

\begin{figure}
\vspace{3.3in}
\includegraphics{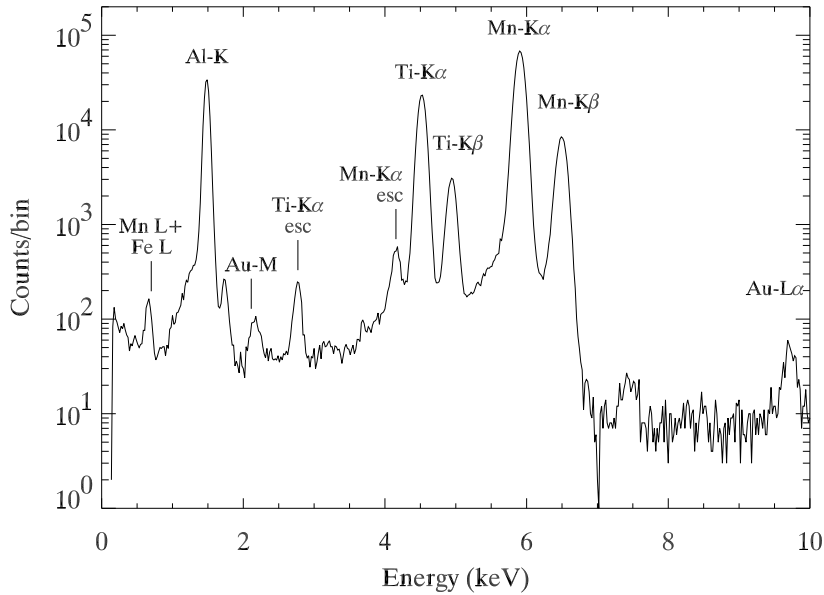}
\caption{The spectrum of the ACIS calibration source on the ACIS-I3 CCD.  Only data from rows 1-256 are included to minimize the CTI degradation.  The spectral lines used in this study are labeled.}
\label{fig:ecsspectrum}
\end{figure}

All the results shown here are based on data taken of the ACIS External Calibration Source (ECS) from a single node of the front-illuminated ACIS-I3 CCD.  Since the discovery of the initial radiation damage, a continuing series of observations have been undertaken just before and after the instruments are safed for perigee passage to monitor the performance of the ACIS CCDs.  ACIS is placed in the stowed position exposing the CCDs to the ECS which produces many spectral features, the strongest of which are Mn-K$\alpha$ (5.9~keV), Ti-K$\alpha$ (4.5~keV), and Al-K (1.5~keV).  Figure~\ref{fig:ecsspectrum} shows a spectrum of the ECS and labels the features used in this study.  The data are taken in the standard Timed Exposure mode with a 3.2~second frame time.  Typical exposure times for each observation range from 5.5 to 8 ksecs.  Approximately 4.5 million events were used in this study, representing about 200~ksec taken during the year 2000.  ACIS events are recorded as 3 x 3 pixel event islands.  The pulseheight of the event is a sum of all the pixels in the island above the split threshold.

\section{CHARACTERISTICS OF ACIS RADIATION DAMAGE} 
\label{sect:raddamage}

A symptom of radiation damage in CCDs is an increase in the number of charge traps.  When charge is transfered across the CCD to the readout, some portion can be captured by the traps and gradually re-emitted.  If the original charge packet has been transfered away before the traps re-emit, the captured charge is ``lost'' to the charge packet.  This process is quantified as charge transfer inefficiency (CTI), the fractional charge loss per pixel.  The front-illuminated ACIS CCDs suffer from radiation damage due to low energy protons.  The framestore covers are thick enough to stop this radiation, so damage is limited to the imaging area of these frame-transfer CCDs.  The parallel CTI at 5.9~keV of the ACIS FI CCDs varies across the focal plane from $1 - 2 \times 10^{-4}$ at the nominal operating temperature of --120$^\circ$~C.

\begin{figure}
\vspace{3.3in}
\includegraphics{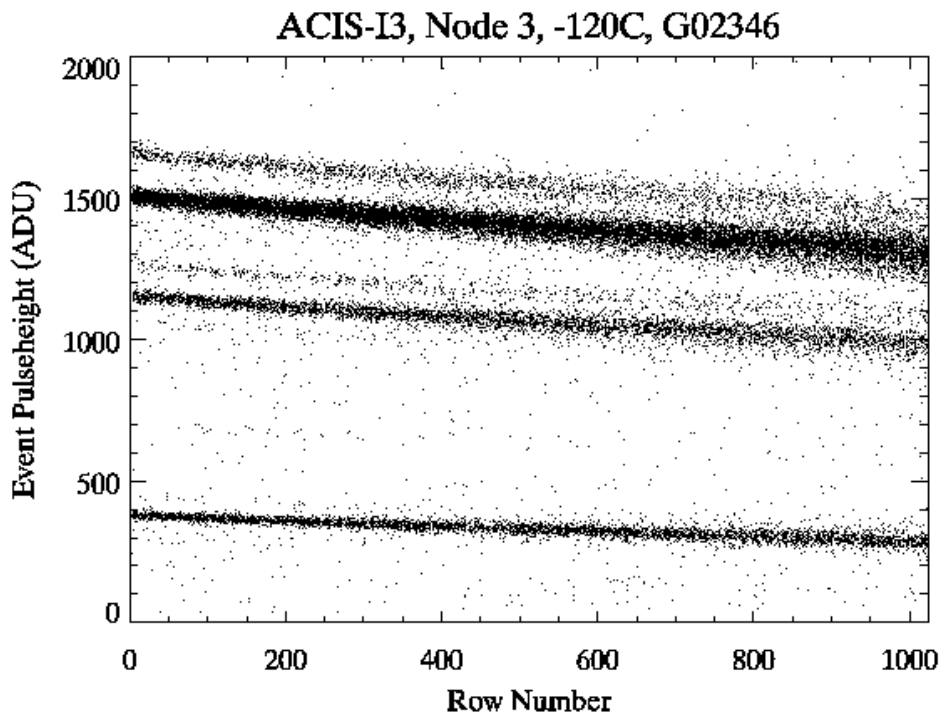}
\caption{A scatter plot of the pulseheight of each X-ray event versus its row number.  The diagonal lines are spectral features in the ACIS calibration source.}
\label{fig:phvrow}
\end{figure}

There are a number of ways in which CTI changes the overall instrument performance.  The cumulative charge loss causes a strong position dependence in the pulseheight, the line response function and the quantum efficiency.  At the bottom of the CCD closest to the readout the performance is the same as the original undamaged CCD.  Figure~\ref{fig:phvrow} is a scatter plot of the pulseheight of each X-ray event versus its row number.  Each of the diagonal lines is a spectral feature in the ACIS calibration source.  The pulseheight of each spectral line decreases with increasing row number due to the increasing transfer distance.  The width of each spectral feature also increases with increasing transfer distance due to a number of factors.  The charge trapping and re-emission process is stochastic, and this additional noise in the pulseheight distribution increases with each transfer.  The trap distribution is non-uniform, so without calibration on the scale of a few pixels, this adds to the spectral width.  Finally, there are variations in trap occupancy which increase the charge loss distribution\cite{saccharge}.  All of these effects are applied to the charge in each pixel, so multi-pixel events will be more degraded than single-pixel events.

\begin{figure}
\vspace{3.3in}
\includegraphics{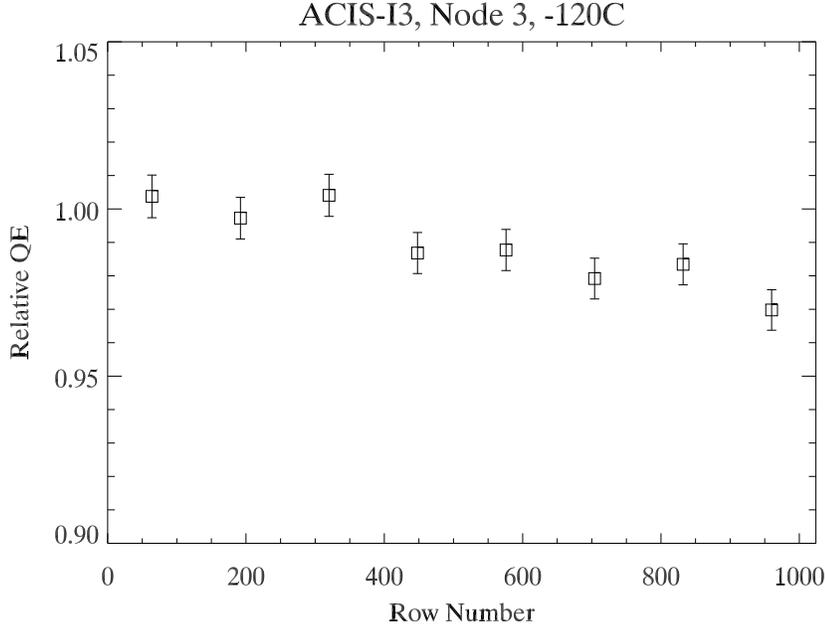}
\caption{The relative quantum efficiency as a function of row number for 5.9~keV X-rays.  The quantum efficiency drops with increasing transfer distance.}
\label{fig:qevrow}
\end{figure}

The charge loss and re-emission process can also change the morphology of the X-ray event.  ACIS events are recorded as 3 by 3 pixel islands.  Each event island is assigned a grade based on the charge distribution.  In this manner X-ray events can be distinguished from the much more common cosmic ray events which have distinctly different event island distributions.  Events with grades that are least likely to be from X-rays are ignored in analysis and a subset are discarded on-board to reduce telemetry requirements.  After the cumulative charge loss and re-emission during transfer, X-ray events can be smeared into morphologies commonly associated with cosmic rays.  The produces a row-dependent loss of quantum efficiency as shown in Figure~\ref{fig:qevrow}.  At the current damage level of the ACIS FI CCDs, the quantum efficiency loss is small and is only important for high energy X-rays which have a higher proportion of multi-pixel events.

\section{CTI CORRECTION} 
\label{sect:cticorr}

If the charge loss process can be modeled, it should be possible to apply a post-facto correction algorithm to reconstruct the original X-ray event and recover some of the detector's spectral resolution.  This type of correction was first shown to be effective for ACIS CCDs by Townsley and collaborators\cite{Townsley00,Townsley02}.  The algorithm described in this paper, which has been implemented by the Chandra X-ray Center as part of the \verb+acis_process_events+ tool,\footnote{For further information on Chandra data analysis: http://cxc.harvard.edu} was inspired by and shares some characteristics with their previous work.

The code is iterative for each event.  A guess is proposed for the original event island before it suffered any corruption from CTI.  The charge loss and charge trailing model is applied to the guess to predict what would be read out from the instrument.  This prediction is compared to the actual event and the deviations are fed back into the initial guess for the next iteration.  Once the differences between the predicted and actual events falls below a certain threshold, the output is the hypothesized uncorrupted event.

\subsection{Charge Loss}

Charge loss in this algorithm is the product of the transfer distance multiplied by the density of charge traps multiplied by the volume of the charge packet.  The volume of the charge packet, and thus the energy dependence of the charge loss, is parameterized as a power law function of energy.  The total charge loss for an isolated pixel is then
$$\delta Q (x,y,Q) = y \, \overline{n}(x,y) \, kQ^\alpha , $$
where $x$, $y$ and $Q$ are the position and charge of the pixel, $\overline{n}$ is the average trap density of pixels $(x,1)$ through $(x,y)$, $k$ is a normalization constant, and $\alpha$ is the power law index.  In this way the position- and energy-dependences are easily separable.

\begin{figure}
\vspace{3.3in}
\includegraphics{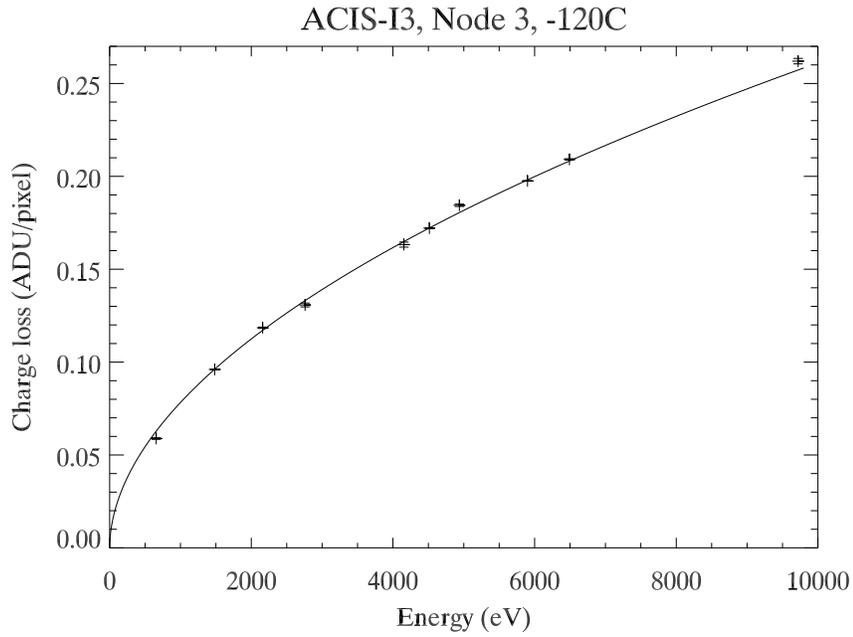}
\caption{This plot shows the charge loss as a function of energy.  Each data point is a fit to a spectral line in the ACIS calibration source.  The solid line is the best fit power law.}
\label{fig:chargeloss}
\end{figure}

We have chosen to parameterize the energy dependence of charge loss, which is related to the volume of the charge packet, as a power law function of the charge with index $\alpha$.  Because it is linked to a fundamental CCD parameter (the size of the charge cloud before the charge is transfered), we have assumed that a single value of $\alpha$ can be used to characterize each CCD.  Figure~\ref{fig:chargeloss} shows the charge loss as a function of energy for one node of the ACIS-I3 CCD along with the best fit power-law.  To calibrate $\alpha$, the charge loss is expressed as a rate (ADU/pixel) and is the linear slope of the pulseheight in the center pixel of the event island as a function of row number for a particular spectral feature.

\begin{figure}
\vspace{3.3in}
\includegraphics{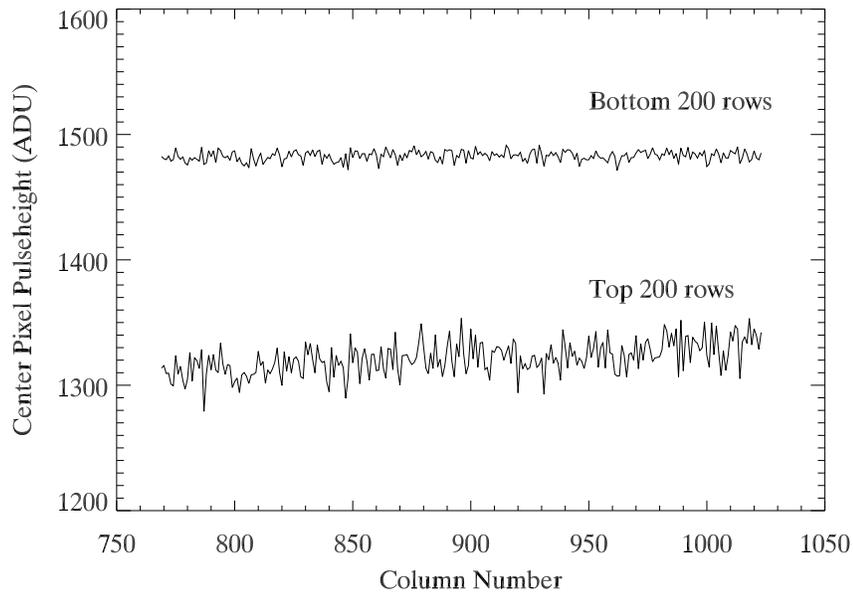}
\caption{The mean center pixel pulseheight in each column for 5.9 keV events.  Data in the bottom 200 rows of the CCD have a much narrower pulseheight distribution than data in the top 200 rows.}
\label{fig:column}
\end{figure}

Also part of the charge loss model is the uniformity of the electron trap distribution across the CCD, $\overline{n}(x,y)$.  There is a known gradient in the radiation damage across the ACIS focal plane, such that the worst CCD has about twice the CTI as the best.  This large-scale gradient needs to be incorporated into the trap distribution map.  In addition, there is non-uniformity on the pixel-scale due to the small number statistics of the traps.  All ACIS observations are dithered over a region of 32 x 32 pixels, so even for point sources, science analysis requires summing over a region at least this large. Figure~\ref{fig:chargeloss} shows that the charge loss in each transfer is much less than one electron for all the relevant X-ray energies.  Even with a completely uniform distribution of charge traps, the variation in the charge loss due to Poisson noise from one column to the next will add significantly to the width of the pulseheight distribution integrated over many pixels.  Figure~\ref{fig:column} shows this small scale effect.  Not only is the mean pulseheight lower at the top of the CCD, but there is a larger variance in pulseheight from column to column.  This is from real variations in the charge trap density due to Poisson noise in the trap distribution.

This trap uniformity has been calibrated using the Al-K$\alpha$ line at 1.5 keV by fitting the charge loss in each column separately.  In principle, any energy spectral line could be used since the map should be energy independent.  Variations in the trap density along each column are included, but are much less important.  The product used by the CTI correction software is an image of the CCD with the value of $y \, \overline{n}(x,y) \, k$ for each pixel. (The trap density, $\overline{n}$, and the volume normalization constant, $k$, are degenerate, so they are calibrated together.)  At a given energy, $y \, \overline{n}(x,y) \, k$ is just the measured charge loss at $(x,y)$ divided by $Q^\alpha$.

\subsection{Charge Trailing}

\begin{figure}
\vspace{3.3in}
\includegraphics{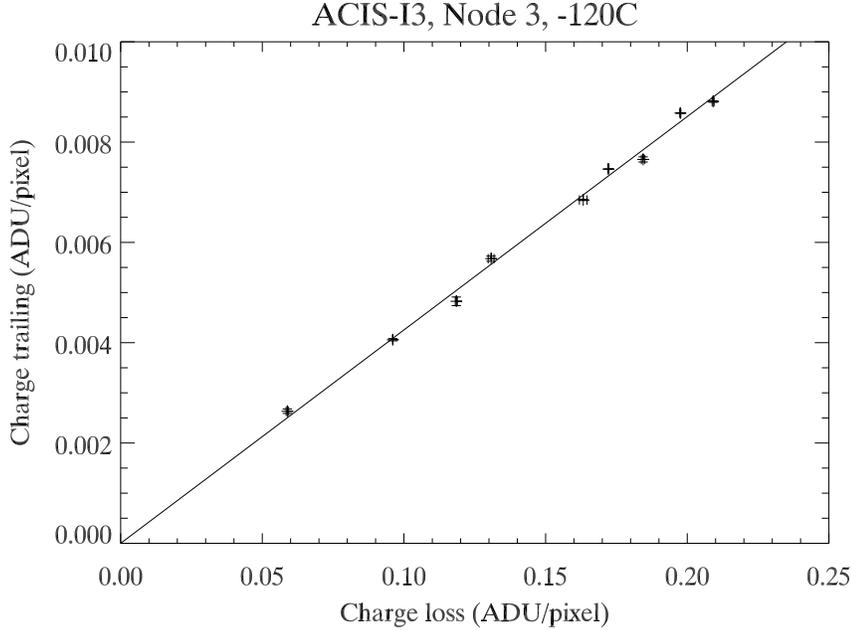}
\caption{This plot shows the charge trailed into the following pixel as a function of the charge loss.  Each data point is a fit to a spectral line in the ACIS calibration source.  The solid line is the best fit linear function with an intercept of zero.}
\label{fig:trailcharge}
\end{figure}

The charge trapping time is assumed to be very short, so that all available traps are filled essentially instantaneously.  As discussed in Section~\ref{sect:raddamage}, the charge that is lost to a particular pixel will be gradually re-emitted.  If the time scale of the re-emission is close to the pixel-to-pixel transfer time (40 $\mu$sec for ACIS), significant charge may be deposited into the pixel immediately following the original charge packet.  About half of the ACIS trap population has time constants in which charge trailing into the following pixel is important\cite{gyp00}.  If this trailed charge is larger than the split event threshold it will be included in the summed pulseheight and it will change the grade assigned to the event.  Figure~\ref{fig:trailcharge} shows the charge trailing as a function of the charge loss.  The charge loss shown here is the same as that in Figure~\ref{fig:chargeloss}.  The charge trailing is expressed as a rate (ADU/pixel) and is the linear slope as a function of row number of the pulseheight in the top pixel of the event island for each spectral feature.  The slope of the linear fit shown in Figure~\ref{fig:trailcharge} is the fraction of the charge loss that is trailed into the following pixel and is used as part of the event reconstruction.

\subsection{Event Island Reconstruction}

The CTI correction model also accounts for sacrificial charge shielding and charge trailing within the event itself.  Thus far we have described the charge loss and trailing from a single isolated event pixel.  Depending on the X-ray energy, ACIS events are often split over two to four pixels.  This causes a more complicated situation in which the charge loss from one pixel can be mitigated by sacrificial charge from a preceding pixel.  Likewise, charge trailing can be reduced by charge in a following pixel.  (For more on sacrificial charge, see Ref.~\citenum{saccharge})

Event island reconstruction begins by calculating the charge loss for each significant pixel in the event assuming an isolated pixel.  Then the pulseheight in each significant pixel is compared to that of the preceding pixel, i.e. with lower row number.  If the preceding pixel has the same or larger pulseheight, no charge is lost from the pixel.  If the preceding pixel has a smaller pulseheight, the effective charge loss is calculated as the difference between the charge loss in each pixel.  In this way, the preceding pixels charge loss has shielded the following pixel from some of the charge traps, thus reducing the charge loss in the following pixel.

The effective charge trailing is calculated in much the same way.  The charge trailed from each significant pixel into the next is calculated assuming a single isolated pixel.  The pulseheight in each significant pixel is compared to that of the following pixel.  If the following pixel has the same or larger pulseheight, no charge is trailed from the pixel into the following pixel.  If the following pixel has a smaller pulseheight, the effective charge trailing is the difference between the charge trailing in each pixel.

\begin{figure}
\vspace{2.5in}
\includegraphics{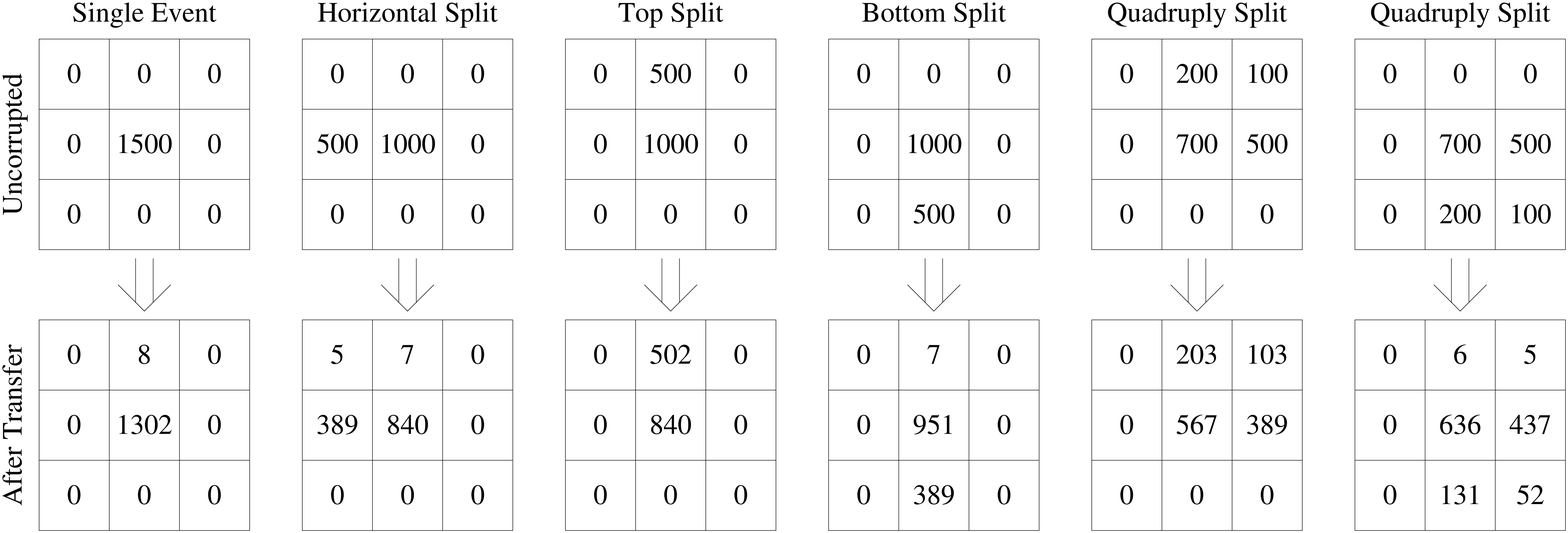}
\caption{Some example 3 x 3 event islands before (top) and after applying our CTI model (bottom).  The transfer direction is downward.  All the original uncorrupted events start with the same total pulseheight of 1500 ADU.}
\label{fig:corrsample}
\end{figure}

The effective charge loss or trailing are then subtracted from or added to the hypothesized uncorrupted event to yield the predicted event which is compared to the actual event received from the instrument.  Figure~\ref{fig:corrsample} shows some example event islands before and after applying our CTI model.  All the sample events have the same original pulseheight, 1500 ADU, or about 6~keV.

\section{PERFORMANCE COMPARISON}
\label{sect:perf}

CTI correction removes the position dependence of pulseheight, provides substantial improvement in spectral resolution and some small improvement in the detection efficiency.  Our primary figure of merit for judging the efficacy of any CTI mitigation approach is the full width at half maximum (FWHM) of spectral features at the top of the CCD (far from the readout) where CTI effects are largest.  On the ACIS-I3 CCD this region corresponds to the aimpoint of the ACIS I-array.  Only ASCA grade 0 (one pixel), 2, 3, 4 (two pixel), and 6 (three or four pixel) events are included.

\begin{figure}
\vspace{3.3in}
\includegraphics{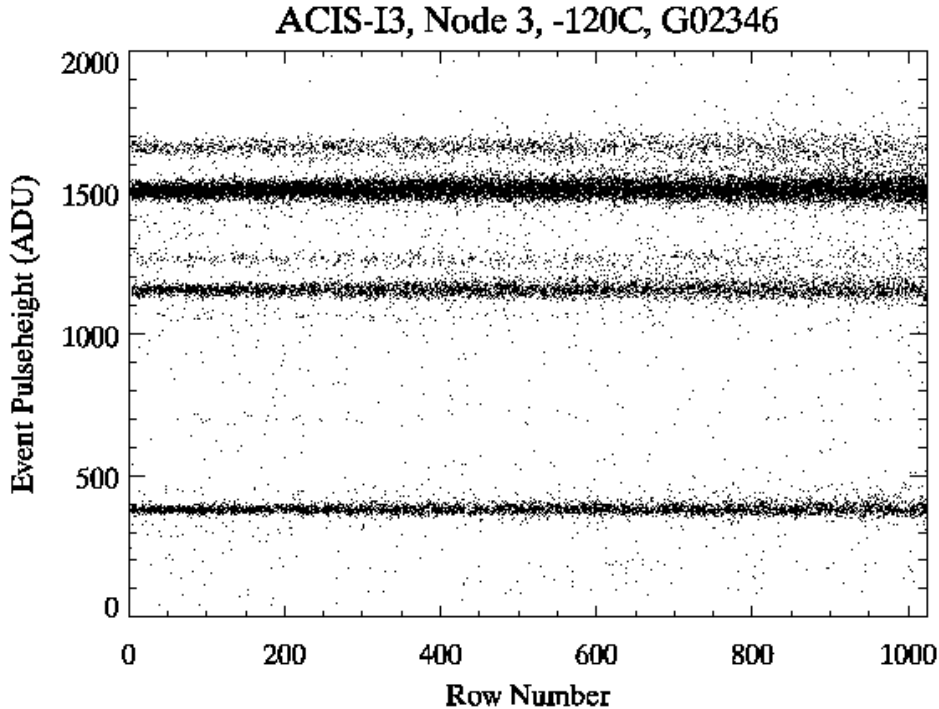}
\caption{A scatter plot of the pulseheight of each X-ray event versus row number after applying CTI correction for comparison with Figure~\ref{fig:phvrow}}
\label{fig:phvrowcorr}
\end{figure}

\begin{figure}
\vspace{4.7in}
\includegraphics{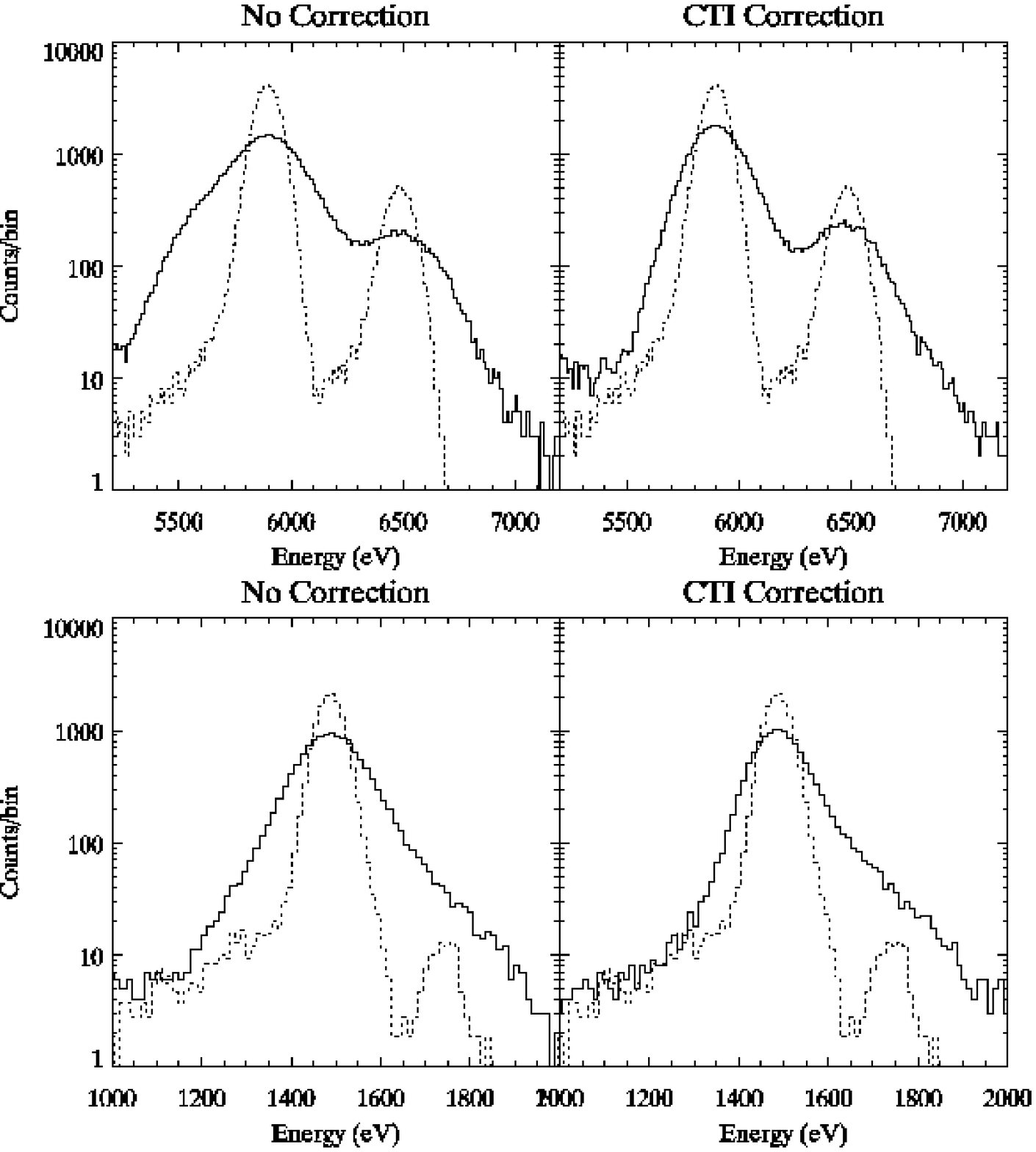}
\caption{Pulseheight spectra around 6 keV (top) and 1.5 keV (bottom) for the top 64-rows of the CCD for uncorrected (left) and CTI corrected (right) data.  The dotted lines show the original undamaged performance.  Note that the y-axis uses log scaling to better show the line shoulders.}
\label{fig:spectra}
\end{figure}

Figure~\ref{fig:phvrowcorr} is a scatter plot of the pulseheight of each X-ray event versus row number after applying CTI correction.  Comparison with Figure~\ref{fig:phvrow} demonstrates that the CTI correction removes the row-dependence of pulseheight and also improves spectral resolution.  Figure~\ref{fig:spectra} shows the pulseheight spectra around 6 and 1.5~keV for the top 64 rows of the CCD where CTI effects are strongest.  Also shown is the original, undamaged performance of the CCD.  The corrected spectra are narrower than the uncorrected, particularly the low energy shoulder.  The high energy shoulder is due to sacrificial charge outside the 3 x 3 event island and cannot be properly corrected with the information in the standard data format\cite{saccharge}.

\begin{figure}
\vspace{3.3in}
\includegraphics{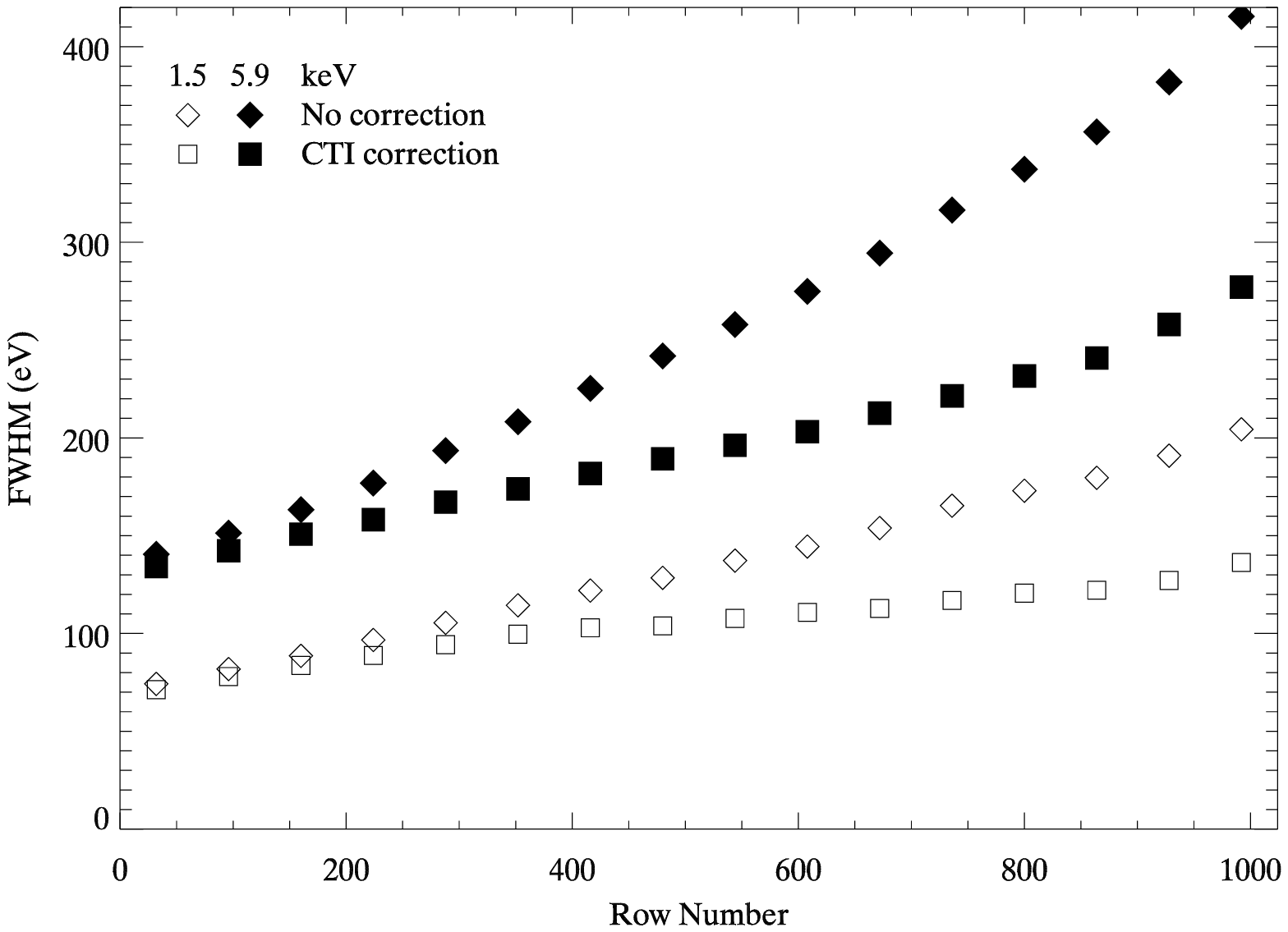}
\caption{FWHM of the spectral lines at 5.9 keV and 1.5 keV as a function of row number for uncorrected and CTI corrected data.}
\label{fig:fitrow}
\end{figure}

Figure~\ref{fig:fitrow} shows the FWHM of the spectral lines at 1.5 and 5.9~keV as a function of row number.  At low row numbers, the FWHM approaches the undamaged value.  Applying the CTI correction improves the FWHM by up to 33\% at both energies.  For the highest 64-row bin, the FWHM at 5.9~keV improves from 415 $\pm$ 1 eV to 277 $\pm$ 1 eV, while the FWHM at 1.5~keV improves from 204 $\pm$ 1 eV to 136 $\pm$ 1 eV.

CTI correction provides a small improvement in detection efficiency by reconstructing event island morphologies so that valid X-ray events are not mistakenly rejected as cosmic ray events.  As shown in Figure~\ref{fig:qevrow}, efficiency for 5.9~keV X-rays drops by about 3\% at high row numbers due to CTI.  A small fraction of the lost events, up to 10\%, is recovered by CTI correction.  One plausible reason for the small size of the improvement is that a sizable fraction of the lost events were in the grade set that is rejected on-board the spacecraft and never telemetered to the ground.  This particular grade set was determined before the radiation damage and at that time consisted of grades that made up 90\% of the background events.  With the advent of CTI, some valid X-ray events can end up in these discarded grades.  Changing the grade reject list to allow these grades would greatly reduce the ACIS background rejection efficiency and cause telemetry saturation, so is not considered a viable option.

\section{DISCUSSION}
\label{sect:disc}

\begin{figure}
\vspace{3.3in}
\includegraphics{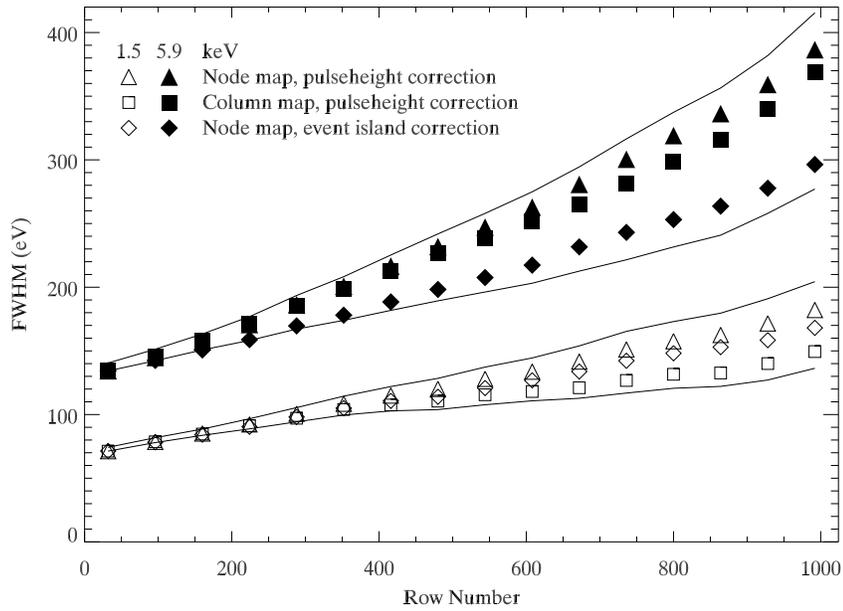}
\caption{FWHM of the spectral lines at 5.9 keV and 1.5 keV as a function of row number for various stages in CTI correction.  The solid lines which bracket the symbols are for uncorrected and fully corrected data.}
\label{fig:fitrow2}
\end{figure}

It is instructive at this point to try and separate out the different elements of the CTI corrector and determine how important they are in improving performance.  Two interesting elements are the small spatial resolution of the charge loss mapping and the detailed correction of each pixel in the event island.  We define four test cases as follows:  i) correcting the total pulseheight of each event using a large-scale charge loss map calibrated for an entire CCD node, ii) correcting the total pulseheight using the standard pixel-scale charge loss map, iii) correcting the event island using the large-scale charge loss map, and iv) the full correction of the event island using the pixel-scale charge loss map.  The FWHM as a function of row number for each of these cases is shown in Figure~\ref{fig:fitrow2}. 

At high energies the largest decrease in FWHM is due to correcting the entire event island (29\% and 33\%) rather than just the summed pulseheight (7\% and 8\%).  Using the more detailed charge loss map only improves the correction by a few percent.  The reverse is true at low energies.  Using the pixel-scale charge loss map and correcting either the summed pulseheight or the event island produces a larger decrease in FWHM (27\% and 33\%) than using the large-scale charge loss map (11\% and 18\%).  The primary difference between low and high energy events is the importance of events split over multiple pixels; at 1.5~keV 79\% of the events are single pixel, 19\% are doubly split and 2\% are multiply split, while at 5.9~keV 47\% are singly split, 33\% are doubly split, and 20\% are multiply split.

This CTI corrector does not, as currently coded, make any correction for CTI in the framestore or serial arrays.  This is appropriate for the ACIS FI CCDs, but not for the ACIS BI CCDs which have substantial CTI in the framestore and serial arrays from manufacturing defects.  In principle, our methodology should be applicable and the Townsley CTI corrector\cite{Townsley00,Townsley02} does correct BI CCDs.  The change in FWHM will be smaller than that seen for the FI CCDs because of the lower level of CTI degradation.

\begin{figure}
\vspace{3.3in}
\includegraphics{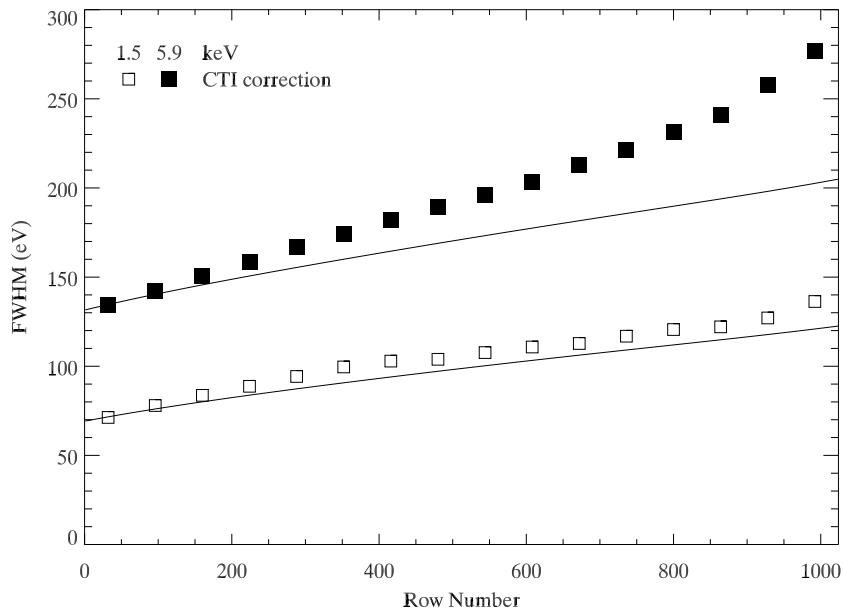}
\caption{FWHM of the spectral lines at 5.9 keV and 1.5 keV as a function of row number for CTI corrected data.  The solid lines are an estimate of the theoretical limit on performance.}
\label{fig:theo}
\end{figure}

Since the charge loss process is stochastic, we can never expect to recover all of the resolution lost to radiation damage through a post-facto software correction.  Figure~\ref{fig:theo} compares the FWHM at 5.9 and 1.5~keV as a function of row number with an estimated theoretical limit on performance.  This limit was determined by assuming that the fluctuations in charge loss are Poissonian and includes the additional effect of split events.  This estimate does not, however, include the effects of sacrificial charge from cosmic ray events.  Since each individual event will have a different sacrificial charge history, the
FWHM will suffer additional degradation.  Ref.~\citenum{saccharge} describes a modification to the basic CTI correction algorithm which provides additional improvement, however since most of the sacrificial charge deposited on the CCD is from cosmic rays that are rejected by on-board event processing, the standard ACIS event lists are insufficient.  Strategies which telemeter additional information about sacrificial charge history with each event are under investigation by the ACIS instrument team.

Some of the additional broadening at 5.9~keV may also be due to the poor calibration of the energy-dependence at low energies.  There are no lines in the calibration source below 700 eV, so the power law energy-dependence is only an extrapolation.  For split events, which are common at high energies, it is quite possible for the split pixels to have pulseheights well below the calibration.  The energy-dependence of charge loss at low energies is currently under investigation.

Clearly, the CTI corrector will require periodic recalibration due to the slow accumulation of additional radiation damage over the mission lifetime.  In addition, since the corrector was calibrated using flight data which includes sacrificial charge from the cosmic rays, changes in the cosmic ray environment will also require recalibration.  At the present time, a full recalibration of the corrector has not been necessary, because the small CTI increase manifests primarily as a change in pulseheight.  This pulseheight change has been more simply corrected by periodic updates to the ACIS gain calibration.  As of July 2004, this functionality is part of the \verb+acis_process_events+ tool.  Based on the current rate of change in CTI and cosmic ray background, an update to the CTI corrector calibration will likely be necessary in the next year or two.

\acknowledgments

We are exceedingly grateful to Leisa Townsley, Pat Broos, and their collaborators at Penn State University for designing the initial ACIS CTI corrector and for many fruitful discussions.  Thanks also go to Gregory Prigozhin and Peter Ford of the ACIS instrument team who were instrumental in our study of ACIS radiation damage.  We also thank Glenn Allen, Dick Edgar, Paul Plucinsky and Chandra X-ray Center for implementing this algorithm in the Chandra pipeline.  This work was supported by NASA contracts NAS 8-37716 and NAS 8-38252.

\bibliography{paper}
\bibliographystyle{spiebib}

\end{document}